\documentclass[trackchanges,twocolumn]{aastex701}

\usepackage{xcolor}
\def\mean#1{\left< #1 \right>}

\def\blue{\color{black}}

\usepackage{amsmath}

\begin{document}


\title{Probing Minihalos in the Late Dark Ages through Velocity Acoustic Oscillations in the 21-cm Power Spectrum
 }


\correspondingauthor{Bin Yue}
\email{yuebin@nao.cas.cn}

\author[0009-0009-6070-0539]{Yan Dong}
\affiliation{State Key Laboratory of Radio Astronomy and Technology, National Astronomical Observatories, Chinese Academy of Sciences, 20A Datun
Road, Chaoyang District, Beijing 100101, China}
\affiliation{School of Astronomy and Space Science, University of Chinese Academy of Sciences, No.1 Yanqihu East Rd, Huairou District, Beijing 101408, China}
\email{dongyan@bao.ac.cn}

\author[0000-0002-7829-1181]{Bin Yue}
\affiliation{State Key Laboratory of Radio Astronomy and Technology, National Astronomical Observatories,  Chinese Academy of Sciences, 20A Datun Road, Chaoyang District, Beijing 100101, China}
\email{yuebin@nao.cas.cn}

\author[0009-0005-1288-8873]{Chen Zhang}
\affiliation{State Key Laboratory of Radio Astronomy and Technology, National Astronomical Observatories, Chinese Academy of Sciences, 20A Datun
Road, Chaoyang District, Beijing 100101, China}
\affiliation{School of Astronomy and Space Science, University of Chinese Academy of Sciences, No.1 Yanqihu East Rd, Huairou District, Beijing 101408, China}
\email{chenzhang@bao.ac.cn}

\author[0000-0003-2579-7039]{Yuewei Wen}
\affiliation{State Key Laboratory of Radio Astronomy and Technology, National Astronomical Observatories, Chinese Academy of Sciences, 20A Datun
Road, Chaoyang District, Beijing 100101, China}
\email{wenyw@bao.ac.cn}
 
\author[0000-0003-3224-4125]{Yidong Xu}
\affiliation{State Key Laboratory of Radio Astronomy and Technology, National Astronomical Observatories,  Chinese Academy of Sciences, 20A Datun Road, Chaoyang District, Beijing 100101, China}
\email{xuyd@nao.cas.cn}

\author[0000-0002-0966-8598]{Ayodeji Ibitoye}
\affiliation{Department of Physics, Guangdong Technion – Israel Institute of Technology, Shantou, Guangdong 515063, People’s Republic of China}
\affiliation{Centre for Space Research, North-West University, Potchefstroom 2520, South Africa} 
\affiliation{Department of Physics and Electronics, Adekunle Ajasin University, P. M. B. 001, Akungba-Akoko, Ondo State, Nigeria}
\email{ayodeji.ibitoye@gtiit.edu.cn}

\author[0000-0001-6475-8863]{Xuelei Chen}
\affiliation{State Key Laboratory of Radio Astronomy and Technology, National Astronomical Observatories, Chinese Academy of Sciences, 20A Datun
Road, Chaoyang District, Beijing 100101, China}
\affiliation{School of Astronomy and Space Science, University of Chinese Academy of Sciences, No.1 Yanqihu East Rd, Huairou District, Beijing 101408, China}
\email{xuelei@bao.ac.cn}

\shorttitle{Minihalos, Dark Ages \& VAOs}
\shortauthors{Dong et al.}
\begin{abstract}
In the Dark Ages, minihalos above the Jeans mass accrete gas from the intergalactic medium (IGM). The accreted gas would be heated to virial temperature and produces 21 cm signal in emission,
{\blue while the majority of gas retained in the IGM is cold and has 21 cm signal in absorption. 
The 21 cm signal of the IGM generally dominates over that of minihalos, 
}
making it difficult to extract the contribution of minihalos from the detected 21 cm signal. 
However, in the late Dark Ages, in regions where the gas has not yet been influenced by the radiation from Pop III stars, the 21 cm signal from the IGM is weak because collisional coupling
has become less efficient. 
More importantly, the relic supersonic dark matter-baryon relative streaming velocities suppress the gas accretion of smaller minihalos and encode their large-scale coherent features into the 21 cm signal, generating the velocity acoustic oscillation (VAO) features on the 21 cm power spectrum. This offers a promising avenue for detecting minihalos. 
We found that, at $z\sim 15$ ($\sim 20$), the first peak of the net VAO wiggle at $k=0.05$ Mpc$^{-1}$ is $\sim 1.4\times 10^{-3}$ ($\sim 3.1\times 10^{-4}$) mK$^2$. 
With a total of 24 lunar nights/8064 hr (60 lunar nights/20160 hr) of integration, the signal from $z\sim 15$ ($\sim 20$) is detectable for an array composed of 82944 (278784) dipoles, spread over $\sim 3.5$ ($\sim 6.3$) km. If the small-scale matter power spectrum is enhanced and minihalos are much more numerous than in the standard $\Lambda$CDM model, the signal is detectable for an even smaller scale array. 
\end{abstract}




\section{Introduction}

The Dark Ages offer an ideal opportunity for probing the primordial fluctuations and dark matter (DM) properties, as the cosmological information has not yet been contaminated by radiation from astrophysical luminous objects. 
{\blue
The redshifted 21 cm line provides the only known direct tomographic
probe of the neutral hydrogen during the Dark Ages (see reviews e.g. \citealt{Furlanetto_2006PhR...433..181F,
Pritchard_2012RPPh...75h6901P,
Burns_2019BAAS...51c...6B,
Furlnetto_2019arXiv190306212F}). 
In addition, there are other potential observables, include the secondary CMB anisotropies from scattering by moving primordial molecular clouds
\citep{1994ApJ...425..372M,2008A&A...490..521S}, the CMB spectrum distortions arising from the transition lines of the first molecules \citep{2025A&A...698A..32K}, as well as the high-energy radiation of DM annihilation or decay \citep{2004PhRvD..70d3502C}. These observables provide indirect or redshift-integrated constraints on the
matter distribution during the Dark Ages; whereas the 21 cm signal can directly
map their three-dimensional distribution and redshift evolution.
} 
{\blue The 21 cm signal from the Dark Ages} can probe small-scale matter density fluctuations  \citep{2004PhRvL..92u1301L,Lewis_2007PhRvD..76h3005L,Furugori_2020MNRAS.494.4334F}, primordial non-Gaussianity \citep{2006PhRvL..97z1301C,2008PhRvD..77j3506C,
Sekiguchi2019JCAP,Floss_2022JCAP...06..020F,Bull2024RSPTA},  primordial magnetic fields \citep{2006MNRAS.372.1060T,Novosyadlyj_2026AA...706A..48N}, 
{\blue 
DM particle properties \citep{Sekiguchi_2014JCAP...08..007S},
}
DM decay/annihilation or interactions \citep{2007MNRAS.377..245V,2014JCAP...11..024E,
Yang_2016EPJP..131..432Y,
Mondal_2024MNRAS.527.1461M,Paul_2026MNRAS.548ag581P}, primordial black holes \citep{Gong_2017JCAP...08..017G,Yang_2021MNRAS.508.5709Y,Zhang2026}, and cosmological models \citep{Mondal_2023NatAs...7.1025M,Okamatsu_2024PhRvL.133m1001O}.

In the absence of radiation from luminous astrophysical objects, 
{\blue
the 21 cm spin temperature is primarily determined by collisional coupling, whereas the gas kinetic temperature is governed by adiabatic cooling, Compton heating, and shock heating.}
Previous investigations have pointed out that, at $z\gtrsim 10$ a significant fraction of the gas can be heated to above the CMB temperature  {\blue by shock heating} \citep{Gnedin_2004ApJ...608..611G,Furlanetto_2004ApJ...611..642F,Furlanetto_2006PhR...433..181F,Jia_2020RAA....20...95J}. 
\citet{Kuhlen2006ApJ} investigated the spin temperature and brightness temperature of the neutral hydrogen in the late Dark Ages ($z\sim 20$) by hydrodynamical simulations and found that collisions are only efficient in minihalos and filaments. Shock heating generates emission signal in these dense regions while the remaining regions still appear in absorption. 
{\blue The spatial contrast in temperature that is created between the over- and under-dense environments provides a favorable opportunity to probe the distribution of neutral hydrogen during this epoch.}
While \citet{Xu2021ApJ} found that such effects can reduce the 21 cm global absorption by up to $\sim 17\%$ in the absence of luminous astrophysical objects.  
\citet{Park_2025NatAs...9.1723P} and \citet{Sikder_2026ApJ..1000L..50S} proposed that the clumping effect that arises from small-scale density fluctuations can change the 21 cm global spectrum and large-scale power spectrum, thus opening a new window  into small-scale structures and DM properties through the global signal or the large-scale fluctuations.

{\blue
Forming from primordial fluctuations and growing continuously during the Dark Ages, the first nonlinear DM structures act as tracers of the primordial fluctuations field.
}
However, such structures are difficult to detect directly, unless the DM particles can annihilate into cosmic rays or photons \citep{Bertone2005PhR,Diemand2005Nature}. 
Once the mass of a DM halo exceeds the Jeans mass \citep{Barkana2001PhR},  
it can accrete gas from the intergalactic medium (IGM). The accreted gas in such a minihalo is finally heated to virial temperature which is generally much  higher than the CMB temperature, then produces the 21 cm signal in emission \citep{Iliev2002APJL,Furlanetto2006}. In contrast, in the Dark Ages the 
IGM temperature is lower than the CMB, so its 21 cm signal is always in absorption.

\citet{Iliev2002APJL,Shapiro_2006ApJ...646..681S,Furlanetto2006} studied the 21 cm emission from minihalos in the Dark Ages.  
{\blue
They found that a single minihalo can reach a high brightness temperature ($\sim$ several hundred mK); however, direct detection is severely limited by its compact angular scale and the blending of multiple minihalos within a single beam
}
Thus, for an interferometer, the received beam-flux is the cumulative contribution from many minihalos. 
{\blue The fluctuations level of the beam-flux}
would then be much lower, just several mK, but still detectable for realistic telescopes.
Their works provide the theoretical basis for using the 21 cm signal to probe the first nonlinear structures.  
{\blue
Nevertheless, in the standard $\Lambda$CDM cosmology, the 21 cm power spectrum of minihalos is generally much weaker than the IGM \citep{Furlanetto2006,Yue2009MNRAS,Sekiguchi2019JCAP}, 
}
because most of the gas resides in the IGM, just a tiny fraction collapses into minihalos \citep{Ciardi_2006MNRAS.366..689C}. 
Therefore, it is difficult to extract the contribution of minihalos from the total 21 cm signal received.

The relative streaming motion between the DM and baryon is generated before recombination and is still supersonic in the Dark Ages. 
\citet{AliHaimoud2014PhRvD} investigated 
{\blue its effects}
on the 21 cm signal from the Dark Ages and found that at large scales, the signal is enhanced by several to tens of percent; at small scales, the signal is suppressed.
Interestingly, at the extremely small scales, the signal is boosted instead. They pointed out that such effects allow us to probe small-scale matter power spectrum by measuring the large-scale signal. 

This relic relative motion can prevent gas from being accreted into smaller minihalos. As a result, the gas content of minihalos is modulated by the relative velocities. As the spatial correlation of the relic DM-baryon relative streaming velocities has oscillatory features \citep{Tseliakhovich2010,Tseliakhovich2011MNRAS,Dalal2010JCAP,AliHaimoud2014PhRvD,McQuinnOLeary2012ApJ,Fialkov2014IJMPD}, all physical processes relevant to the gas content in minihalos {\blue must also have similar oscillatory features. These are the velocity acoustic oscillations (VAO) features.

During the Cosmic Dawn, the formation of Pop III stars is modulated by the DM-baryon relative motion \citep{Stacy2011ApJL,Greif2011ApJ,Naoz2013,Fialkov2012MNRAS.424.1335F,Visbal2012Nature,Kulkarni2021ApJ,Schauer2021MNRAS}.  
The 21 cm signal of the IGM influenced by the Ly$\alpha$ and X-ray radiation from Pop III stars therefore exhibits oscillations. 
The earliest studies of this oscillatory imprint in the 21 cm signal
were presented by \citet{Dalal2010JCAP} and
\citet{Visbal2012Nature}, subsequently followed  by \citet{McQuinnOLeary2012ApJ,Fialkov2013MNRAS,Mirocha2018MNRAS,Munoz2019,Hotinli2022PhRvD,SarkarKovetz2023PhRvD,Hayato2023,Zhang2024,Zhang2026,Du2026} and so on, in a variety of astrophysical and cosmological settings. These studies showed that the large-scale acoustic structures of the DM-baryon relative velocity field can generate prominent oscillations on the
21 cm power spectrum in the Cosmic Dawn. 
} 
 
{\blue
In the Dark Ages, both minihalos and the IGM have the 21 cm signal. For minihalos the signal is in emission, while for the IGM the signal is in absorption before it is heated to above the CMB. However, it is hard to separate minihalos and the IGM in the received total signal, because their signals have similar large-scale power spectrum shape and the IGM always dominates over the total signal. The DM-baryon relative motion provides us a chance to detect the contributions from minihalos in the received total signal. Since the gas content of minihalos is reduced by the relative motion \citep{Naoz2013}, their 21 cm signal must also have the oscillatory features. Such features remain largely unexplored in literature. If the 21 cm signal of the IGM is not or at most just very slightly influenced by this relative motion, then the VAO features provide us with unique information about minihalos through the 21 cm signal.
}
This is particularly important for the late Dark Ages ($z\sim 30 -20$, sometimes may extend to $z\sim 15$ if the IGM has not yet been influenced by radiation from luminous objects) when minihalos are still small enough and more sensitive to the relative velocities, while the 21 cm signal of the IGM becomes weak because the collision-coupling becomes less efficient \citep{Barkana2001PhR,Furlanetto_2006PhR...433..181F,Pritchard_2012RPPh...75h6901P}.
Therefore, the late Dark Ages is a golden epoch for probing minihalos via the VAO features.

{\blue
Regarding the IGM, in the presence of Pop III stars whose formation is modulated by the relative motion, its 21 cm signal would be influenced by the relative motion as well, through the Ly$\alpha$ and X-ray radiation from these stars (e.g. \citealt{Dalal2010JCAP,Visbal2012Nature}). However, in the Dark Ages when Pop III stars have not yet formed, it remains unclear whether the 21 cm signal of the IGM also has oscillatory features.
It relies on whether the shock heating---the dominant mechanism capable of changing the gas kinetic temperature---works efficiently in the IGM, because the shock heating is modulated by relative motion \citep{Oleary2012ApJ}. 
Many papers \citep{Gnedin_2004ApJ...608..611G, McQuinnOLeary2012ApJ,Xu2021ApJ,Jia_2020RAA....20...95J,Park_2025NatAs...9.1723P} found that the shock heating can change the gas temperature significantly, but they did not analyze the IGM and halos separately.  \citet{Furlanetto_2004ApJ...611..642F} pointed out that even at $z\sim 20$, a significant fraction of the IGM can be heated to $\gtrsim 10^4$ K by shocks. If this is true, then the 21 cm signal of the IGM would also exhibit oscillatory features. However their work was based on analytical model, therefore probably  not accurate for this problem. \citet{Shapiro_2006ApJ...646..681S} also analyzed the 21 cm signal of the IGM and minihalos separately in their hydrodynamical simulations. Their results show that at $z\sim 20$ the 21 cm signal of the IGM is basically in absorption, and is unlikely to be heated significantly. 
Neither of these two works incorporated the DM-baryon relative motion. 
Moreover, \citet{Oleary2012ApJ} simulated the relative motion-induced shock heating and found that it is confined to the immediate vicinities of halos, whereas the IGM in a large volume is not significantly heated. Based on this, \citet{McQuinnOLeary2012ApJ} found that the temperature fluctuations raised by the relative motion-induced shock heating should be $\lesssim 10\%$.  Although not explicitly illustrated in the paper, it can be inferred from their results that in the Dark Ages, the IGM is unlikely to exhibit prominent VAO features in the 21 cm power spectrum. 
In addition to calculating the VAO signals from minihalos in the late Dark Ages, in this paper we will also evaluate the 21 cm signal from the diffuse IGM in the presence of DM-baryon relative motion, and
aim to quantify whether the IGM will contaminate the VAO features of minihalos, via hybrid of hydrodynamic simulations and semi-numerical simulations.}

The redshifted 21 cm signal from the Dark Ages falls in a very low-frequency band, where terrestrial ionospheric absorption/distortion and radio-frequency interference are severe. The lunar far-side is therefore one of the best locations for such observations. A number of lunar-based or lunar-orbit projects/concepts have been proposed, including DAPPER \citep{Burns2021DAPPER}, LuSEE-Night \citep{Bale2023LuSEE}, DSL (also known as Hongmeng) \citep{Chen2021DSL}, FARSIDE \citep{Burns2021FARSIDE}, 
LARAF \citep{Chen_2024RSPTA.38230094C}, 
FarView \citep{Polidan2024,Burns_2026arXiv260116170B}, CoDEX \citep{Koopmans_2021ExA....51.1641K}, and DEX \citep{Brinkerink2025}.
\citet{Bevins_2026ApJ...997L..35B} gave precise predictions of the magnitude and frequency range of the sky-averaged 21 cm signal based on existing observational constraints on cosmology.
\citet{Pober2025ApJ} quantified how foreground wedge avoidance would degrade the sensitivity of the Dark Ages 21 cm power spectrum observations.
\citet{Wen_2026arXiv260529276W} investigated the feasibility of designing a lunar far-side interferometer array for probing the small-scale 21 cm power spectrum and constraining cosmological parameters.
\citet{Zhang_2026arXiv260623314Z}
developed a wedge-aware Fisher framework to estimate the influence of foreground wedge-avoidance on constraining the primordial non-Gaussianity.

In this paper, we investigate the VAO features on the 21 cm power spectrum of minihalos in the late Dark Ages, and estimate their detectability for future lunar-based low frequency arrays. The paper is organized as follows. In Sec. \ref{sec:individual}, we introduce the 21 cm brightness temperature of individual minihalos. In Sec. \ref{sec:VAO-21cm}, we calculate the VAO wiggles on the 21 cm power spectrum of minihalos in the late Dark Ages. In Sec. \ref{sec:detectability}, we estimate the detectability of this signal using the lunar-based low-frequency array. In Sec. \ref{sec:contamination}, we investigate whether the the 21 cm signal of the IGM will also be influenced by the DM-baryon relative streaming velocities and whether it will contaminate the signal from minihalos. 
In Sec. \ref{sec:discussion} we discuss uncertainties relevant to the lightcone effect and the foreground.
Finally in Sec. \ref{sec:summary}  we  give our summary.
{\blue Throughout this paper, we adopt the Planck 2018 cosmology parameters for $\Lambda$CDM model, $\Omega_{\rm m}=0.3153$, $\Omega_{\rm \Lambda}=0.6847$, $\Omega_{\rm b}=0.0493$, $h=0.6736$,  $\sigma_8=0.8111$, and $n_{\rm s}=0.9649$ \citep{Plank2018}. }

\section{The 21 cm signal of an individual minihalo}\label{sec:individual}

In the late Dark Ages, the gas temperature of a minihalo is generally much larger than the CMB temperature, so at the 21 cm frequency it acts as a bright spot. \citet{Iliev2002APJL,Shapiro_2006ApJ...646..681S} has developed a model for calculating the brightness temperature of such a point source by assuming that its gas profile follows the truncated isothermal sphere (TIS) model \citep{Shapiro1999MNRAS,IlievShapiro2001}. Given the gas density profile $\rho_{\rm gas}(r)$ and temperature profile $T_{\rm K}(r)$, the spin temperature profile is \citep{Furlanetto2007MNRAS}
\begin{equation}
T^{-1}_{\rm S}(r)=\frac{ T^{-1}_{\rm CMB}+x_c(r) T^{-1}_{\rm K}(r)+x_\alpha(r)T^{-1}_\alpha(r)}{1+x_c(r)+x_\alpha(r)},
\label{eq:T_S}
\end{equation}
where $T_{\rm CMB}$ is the CMB temperature at redshift $z$; $x_c$ is the collisional coefficient which depends on gas density and temperature \citep{Zygelman2005ApJ,Furlanetto2007MNRAS}; $x_\alpha$ is the coefficient of hydrogen-Ly$\alpha$ scattering (the Wouthuysen–Field effect, \citet{Wouthuysen_1952AJ.....57R..31W,Field_1959ApJ...129..536F} ). Throughout this paper, we limit our investigation to the Dark Ages where there is no radiation field other than the CMB, so we ignore the $x_\alpha$ term in Eq. (\ref{eq:T_S}). Moreover, since gas in the virialized halo is dense (the overdensity $\delta \gtrsim$ several tens) and optically thick to Ly$\alpha$ photons, even in the presence of Ly$\alpha$ radiation, the coupling is still dominated by collisional processes.

In the TIS model, $T_{\rm K}$ is independent of $r$ and is set to be the virial temperature. Following \citet{Shapiro1999MNRAS,IlievShapiro2001}, the virial temperature of a halo with mass $M_{\rm h}$ is 
\begin{equation}
T_{\rm{vir}} = 7.843 \times 10^5 \left( \frac{\mu}{0.59} \right) \left( \frac{M_{\rm h}}{10^{12} M_{\odot}} \right)^{2/3} F(z) h^{2/3} \, ~~~\mathrm{K},
\label{T_vir}
\end{equation}
where $\mu$ is the mean molecular weight, which we set to $\mu=1.22$ for the neutral gas in minihalos, and $h$ is the dimensionless Hubble constant, and
\begin{equation}
F\!\left(z\right)
=
\left[
\frac{\Omega_{\mathrm{m}}}{\Omega^z_{\rm m}\!\left(z\right)}
\frac{\Delta_{\mathrm{c}}(z)}{18\pi^{2}}
\right]^{1/3}
\left(1+z\right),
\label{F_z}
\end{equation}
in which $\Omega_{\rm m}$ is the matter density parameter at present. The quantity 
\begin{equation}
\Delta_{\mathrm{c}}(z) = 18\pi^2 + 82(\Omega^z_{\mathrm{m}}(z) - 1) - 39(\Omega^z_{\mathrm{m}}(z) - 1)^2,
\end{equation}
is the nonlinear virial overdensity with respect to
the critical density at the collapse redshift \citep{Bryan1998ApJ} and
\begin{equation}
{\blue
\Omega^z_{\mathrm{m}}(z)=\frac{\Omega_{\rm m}(1+z)^3}{\Omega_{\rm m}(1+z)^3+\Omega_{\Lambda}},
}
\end{equation}
in which $\Omega_\Lambda$ is the dark energy density parameter today.

In the TIS model, the gas density profile of a halo is 
\begin{equation}
    \rho_{\rm gas}(r,M_{\rm h}, z)
    =  \frac{\Omega_{\rm b}}{\Omega_{\rm m}}\rho_0(z)
    \left[
    \frac{21.38}{9.08 + \left( r/r_0 \right)^2}
    - \frac{19.81}{14.62 + \left( r/r_0 \right)^2}
    \right],
\end{equation}
where $\Omega_{\rm b}$ is the baryon density parameter, and 
\begin{equation}
{\blue
    \rho_0(z) = 3.382 \times 10^{-25} [F(z)]^3 h^2
    ~{\rm g~cm^{-3}},
    }
\label{rho_0}
\end{equation} is the TIS normalization density. Following the standard TIS model, we identify the halo radius with the
truncation radius,
\begin{equation}
    R_{\rm h}\equiv r_t
    =
    \left[
    \frac{3M_{\rm h}}
    {4\pi \Delta_{\rm c}(z)\rho_{\rm crit}(z)}
    \right]^{1/3},
\end{equation}
and define the TIS core radius by
\begin{equation}
    r_0=\frac{R_{\rm h}}{\zeta_t},
\end{equation}
with $\zeta_t\simeq30.04$ \citep{Shapiro1999MNRAS,IlievShapiro2001}.
{\blue
The spin-temperature profile $T_{\rm S}(r)$ 
is obtained by substituting the above $T_{\rm vir}$ and $\rho_{\rm gas}$ into
Eq. (\ref{eq:T_S}).
}

\begin{figure}
    \centering
    \includegraphics[width=0.8\linewidth]{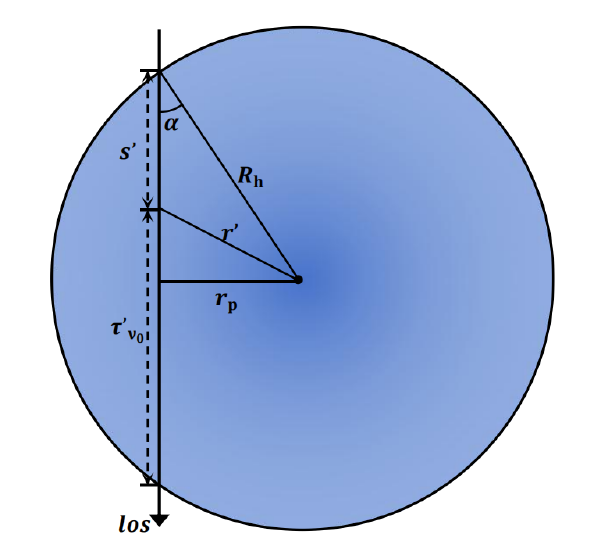}
    \caption{The geometry of the 21 cm brightness temperature for a line of sight with projected distance $r_{\rm p}$ to the center of the minihalo.
 }
    \label{fig:T_b_nu0}
\end{figure}

The brightness temperature at the rest-frame 21 cm transition frequency $\nu_0=1420.4057\,{\rm MHz}$ for a line of sight with projected distance $r_{\rm p}$ from the minihalo center is
\begin{equation}
{\blue
T_{\rm b, \nu_0}(r_{\rm p})=T_{\rm CMB}e^{-\tau_{\nu_0}}+
\int_0^{2\sqrt{R_{\rm h}^2-r_{\rm p}^2}}
T_S(r') e^{-\tau'_{\nu_0}}\kappa_{\nu_0}(r') ds' ,
}
\label{eq:T_b_nu0}
\end{equation}
where the geometry relation between $r'$ and $s'$ is given in Fig. \ref{fig:T_b_nu0}. The optical depth $\tau'_{\nu_0}$ is the integration of the absorption coefficient $\kappa_{\nu_0}$ from $s'$ to $2\sqrt{R_{\rm h}^2-r_{\rm p}^2}$. $\tau_{\nu_0}$ in the first term of the right hand side of Eq. (\ref{eq:T_b_nu0}) is the optical depth for $s'=0$.
The absorption coefficient $\kappa_{\nu_0}(r)$ is
\begin{equation}
\kappa_{\nu_0}(r)=
\frac{3c^2A_{10}n_{\rm HI}(r)}{32\pi\nu_{0}^2}
\frac{T_\star}{T_{\rm S}(r)}\phi(\nu_0),
\end{equation}
where $A_{10}=2.85\times10^{-15}\,{\rm s}^{-1}$ is the spontaneous decay rate of the 21 cm transition, $T_\star=h_{\rm P}\nu_{0}/k_{\rm B}=0.068\,{\rm K}$, and $\phi(\nu_0)$ is the center of the normalized line profile. Following the thermal-Doppler approximation used by \citet{Iliev2002APJL}, we take
\begin{equation}
\phi(\nu_0)=\frac{1}{(1+z)\Delta\nu_{\rm eff}},
\end{equation}
and 
\begin{equation}
\Delta\nu_{\rm eff}(M_{\rm h},z)=
\left(2\pi\mu\right)^{1/2}
\frac{\nu_{0}\sigma_V(M_{\rm h},z)}{c(1+z)} ,
\end{equation}
for which $\sigma_V$ is the one-dimensional velocity dispersion of a TIS halo, c is the light speed. It is calculated from the TIS solution as
\begin{equation}
\sigma_V(M_{\rm h},z)=
\alpha_{\rm TIS}(z)\left[G M_{\rm h}H(z)\right]^{1/3},
\end{equation}
where $G$ is the gravitational constant, $H(z)$ is the Hubble parameter, and
\begin{equation}
\alpha_{\rm TIS}(z)=
(13.5)^{1/6}
\left(\frac{\mathcal{R}_{\rm TIS}}{\zeta_t^2 \alpha_p}\right)^{1/2}
\left[\Delta_{\rm c}(z)\right]^{1/6}.
\end{equation}
Here $\mathcal{R}_{\rm TIS}=530$ and
$\alpha_p=3.68$ are the dimensionless TIS parameters adopted in our calculation.

\begin{figure}
    \centering
    \includegraphics[width=0.8\linewidth]{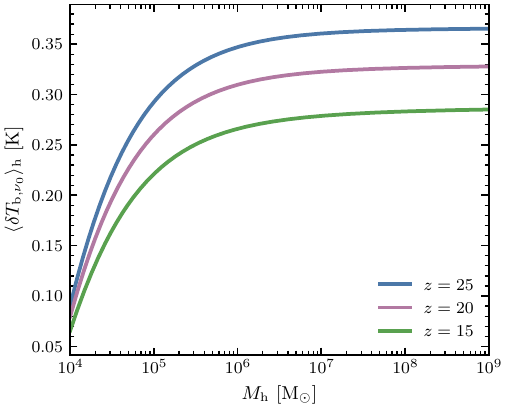}
    \caption{The mean differential brightness temperature at $\nu_0$ for a single minihalo as a function of halo mass  at $z=25$, $20$ and $15$, respectively. 
 }
    \label{fig:deltaTb_vs_Mh}
\end{figure}

Fig. \ref{fig:T_b_nu0} illustrates the geometry for Eq. (\ref{eq:T_b_nu0}). The face-averaged brightness temperature of this halo is 
\begin{equation}
\mean{T_{\rm b,\nu_0}}_{\rm h}=\frac{1}{\pi R^2_{\rm h}}\int_0^{R_{\rm h}} T_{\rm b,\nu_0}(r_{\rm p}) 2\pi r_{\rm p} {\rm d}r_{\rm p} 
\end{equation}
The differential brightness temperature in contrast to the CMB is 
\begin{equation}
\mean{\delta T_{\rm b,\nu_0}}_{\rm h}=\frac{\mean{T_{\rm b,\nu_0}}_{\rm h}-T_{\rm CMB}(z)}{1+z}.
\end{equation}
In Fig. \ref{fig:deltaTb_vs_Mh} we plot the $\mean{\delta T_{\rm b,\nu_0}}_{\rm h}$ as a function of $M_{\rm h}$ at $z=25$, 20 and 15 respectively. The brightness temperature drops rapidly below $\sim 10^4~M_\odot$.

\section{The VAO features in the 21 cm power spectrum of minihalos}\label{sec:VAO-21cm}

\begin{figure}
    \centering
    \includegraphics[width=0.8\linewidth]{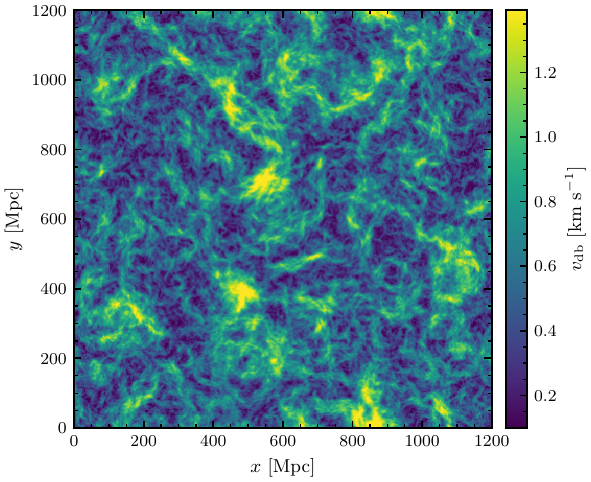}
    \includegraphics[width=0.8\linewidth]{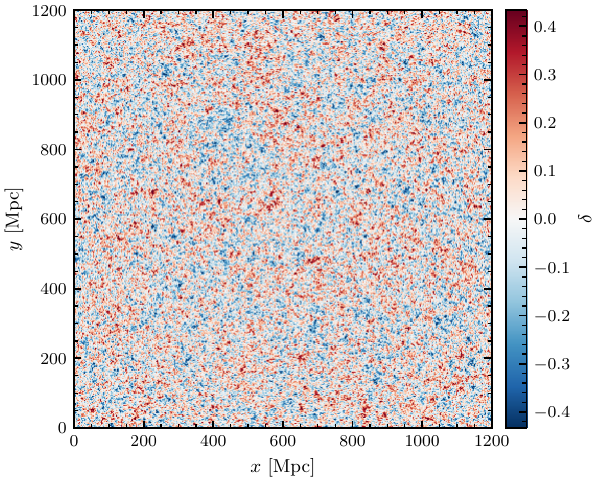}
    \caption{ 
Slices of the DM-baryon relative streaming velocity field ({\it top}) and
matter overdensity field ({\it bottom}) at $z=20$. The slice thickness is
$3\,{\rm Mpc}$.
 }
    \label{fig:vdb}
\end{figure}

\citet{Zhang2024} modified the original {\tt 21cmFAST} \citep{Mesinger2011MNRAS,Murray2020,Qin2020MNRAS,Park2019MNRAS,Munoz2022MNRAS,Davies2025} to involve the effects of the DM-baryon relative streaming velocity on the formation of Pop III stars and treat the Lyman-Werner feedback self-consistently.  
 It has been used to predict the VAO features arising from primordial black holes \citep{Zhang2026} and to investigate the feasibility of constraining H$_2$ cooling and Lyman-Werner feedback parameters in Pop III stars formation \citep{Du2026}.
We use this code to generate the cosmological density field and the DM-baryon relative streaming velocity field.   
We set box length $L_{\rm box}=1200$ Mpc and divide it into 400$^3$ cells. Each cell has a size $x_{\rm cell}=3$ Mpc, 
{\blue
which is consistent with the coherence scale of the DM-baryon relative streaming velocities \citep{Tseliakhovich2010,Fialkov2012MNRAS.424.1335F,Munoz2019}.
} 
{\blue 
Fig. \ref{fig:vdb} shows
}
slices of the density and velocity fields at $z=20$.

In the presence of DM-baryon relative streaming velocity, the gas fraction in smaller minihalos is reduced. \citet{Naoz2013} has investigated this using numerical simulations and has provided a fitting formula for the gas fraction as a function of halo mass and relative velocity $v_{\rm db}$,
\begin{equation}
    f_g(M_{\rm h}, z, v_{\rm db}) = \frac{\Omega_{\rm b}}{\Omega_{\rm m}} \left[ 1 + (2^{\alpha/3} - 1) \left( \frac{M_c(v_{\rm db}, z)}{M_{\rm h}} \right)^\alpha \right]^{-3/\alpha},
\end{equation}
 where the shape parameter for the profile $\alpha=1$, as given by \citet{Gnedin2000ApJ}. The pivotal quantity in this formalism is the characteristic mass, $M_c(v_{\rm db}, z)$, which represents the mass scale  
 {\blue at}
 which the gas content is reduced to half of the cosmic mean fraction.

{\blue
We fit the characteristic mass $M_{\rm c}$ reported by \citet{Naoz2013} as

\begin{equation}
\log(M_{\rm{c}}) = a_0(x_{\rm db}) + a_1(x_{\rm db})\log(1+z) + a_2(x_{\rm db})\log^2(1+z),
\label{eq:logM_c}
\end{equation}
where
\begin{equation}
x_{\rm db}\equiv
\frac{v_{\rm db}}{v^{\rm N13}_{\rm ref}(z)},\qquad
v^{\rm N13}_{\rm ref}(z)\equiv
30\frac{1+z}{1011}\ {\rm km\,s^{-1}}.
\end{equation}
$v^{\rm N13}_{\rm ref}$ is actually the rms of the relative streaming velocities in \citet{Naoz2013}.
}
 
Parameters  $a_0$, $a_1$ and $a_2$ are independent of $z$ and are approximated by,
{\blue
\begin{align}
a_0&= -0.1155 x_{\rm db} + 3.5362,\nonumber \\
a_1 &= 0.9033 x_{\rm db} + 1.2823,\nonumber \\
a_2 &= -0.2985 x_{\rm db} - 0.4199. 
\end{align}
}

{\blue The \citet{Naoz2013} simulations used the WMAP 5-year cosmology parameters \citep{Komatsu_WAMP5_2009ApJS..180..330K}, while we use the Planck 2018 cosmology parameters. In principle there are inconsistencies. However, we believe this will just have minor impacts on our results. We check that in the Planck 2018 cosmology, the rms of the relative streaming velocities $\sigma_{v_{\rm db}}=28.85\,~{\rm km\,s^{-1}}$ at the recombination era, just $\sim$4\% different from the WMAP 5-year cosmology which is 30 km s$^{-1}$. In our calculations, we use the $M_{\rm c}$ in Eq. (\ref{eq:logM_c}) as a sharp cutoff. Minihalos above this characteristic mass are not influenced by $v_{\rm db}$; while minihalos below this mass contain no gas. In such a treatment, the detailed dependence of the $f_{\rm g}$ shape on cosmology is irrelevant to our calculations. The oscillatory features in the 21 cm signal are generated by the spatial variations of $M_{\rm c}$ that arise from the spatial variations of $v_{\rm db}$. We test that, if we vary the $v_{\rm db}$ by $\sim 4\%$ globally, the amplitude of the minihalos VAO wiggles would just change by $\sim 3.6\%$ at $z=20$. 
}

For a cell with density $\delta_{\rm cell}$ and DM-baryon relative streaming velocity $v_{\rm db}^{\rm cell}$, we calculate the 21 cm signal from minihalos in the cell 
\begin{equation}
\begin{aligned}
\delta T^{\rm cell}_{\mathrm{b,MH}}
&= \frac{c(1+z)^4}{\nu_0 H(z)}
\int_{M_{\min}(v^{\rm cell}_{\rm db},z)}^{M_{\max}(z)}
\Delta\nu_{\mathrm{eff}} \,
\mean{\delta T_{\mathrm{b},\nu_0}}_{\rm h} \, A \\
&\quad \times
\frac{dn}{dM_{\mathrm h}}
\left(M_{\mathrm h},z \mid
\delta_{\rm cell},M_0^{\rm cell}\right)
\, dM_{\mathrm h},
\end{aligned}
\end{equation}
where $M_0^{\rm cell}=\rho_{\rm m}x^3_{\rm cell}$ is the mean mass of one cell, $\rho_{\rm m}$ is the mean matter density of the Universe, and $A(M_{\rm h},z)=\pi R_{\rm h}^2$ is the geometric cross section of a minihalo.

For simplicity we set the minimum mass of minihalos sustaining gas in the presence of $v_{\rm db}$ as $M_{\rm min}=M_c$. We set $M_{\rm max}$ as the virial mass corresponding to $T_{\rm vir}=10^4$ K. 
The conditional halo mass function  $dn/dM_{\rm h}$ is given by the Extended Press-Schechter (EPS) scenario \citep{Mo1996},
\begin{equation}
\frac{dn}{dM_{\mathrm h}} = \rho_{\rm m}(1+\delta_{\rm cell})
\sqrt{\frac{1}{2\pi}} \frac{1}{M_{\mathrm h}} \frac{d\sigma^2}{dM_{\mathrm h}} \frac{\delta_1}{\sigma_1^3} \exp \left( -\frac{\delta_1^2}{2\sigma_1^2} \right),
\label{eq:dn_dm}
\end{equation}
in which $\sigma(M_{\rm h})$ is the rms of the density field fluctuations, smoothed at mass scale $M_{\rm h}$, and 
\begin{equation}
\sigma^2_1=\sigma^2(M_{\rm h})-\sigma^2(M_0^{\rm cell}).
\end{equation}
The overdensity
\begin{equation}
\delta_1=\delta_{\rm c}(z)-\delta_{\rm L}(\delta_{\rm cell},z),
\end{equation}
where $\delta_{\rm c}(z)$ and $\delta_{\rm L}$ are linear extrapolations of critical overdensity for halo collapse and $\delta_{\rm cell}$ to present-day \citep{Mo1996} respectively.

Since the volume-averaged EPS abundance does not exactly reproduce the
Sheth-Tormen (ST) mass function \citep{ST_1999MNRAS.308..119S} for the mean Universe, we renormalize the
conditional EPS mass function in Eq. (\ref{eq:dn_dm}) by a factor defined as 
\begin{equation}
{\cal R}_{\rm ST}(M_{\rm h},z)
=
\frac{
dn/dM_{\rm h}|_{\rm ST}
}{
\left\langle
dn/dM_{\rm h}|_{\rm EPS}
\right\rangle
},
\label{eq:st_rescale_factor}
\end{equation}
where the numerator is the ST halo mass function, and the denominator is the average of Eq. (\ref{eq:dn_dm}) taken over all cells in the cosmological box. 
This prescription is quite popular (e.g. \citealt{Mesinger2011MNRAS}), it preserves the cell-to-cell modulation predicted by EPS, while
ensuring that the box-averaged halo abundance is normalized to the ST mass
function.

The 21 cm brightness temperature of the IGM in each cell is \citep{Pritchard_2012RPPh...75h6901P}
\begin{equation}
\begin{aligned}
\delta T^{\rm cell}_{\rm b,IGM}
&= 27 x_{\rm HI}(1+\delta^{\rm IGM}_{\rm cell})
\left(\frac{\Omega_b h^2}{0.023}\right) \\
&\quad \times
\left(\frac{0.15}{\Omega_{\rm m} h^2}
      \frac{1+z}{10}\right)^{1/2}
\left(1-\frac{T_{\rm CMB}}{T^{\rm IGM}_{\rm S}}\right)
\,\mathrm{mK},
\end{aligned}
\label{eq:T_b_IGM}
\end{equation}
where the spin temperature of the IGM, $T_{\rm S}^{\rm IGM}$, is calculated from Eq. (\ref{eq:T_S}), by replacing the halo density and temperature with those of the IGM in that cell. The IGM overdensity satisfies  $1+\delta_{\rm cell}^{\rm IGM}=(1+\delta_{\rm cell})(1-f_{\rm coll}(\delta_{\rm cell},v_{\rm db}^{\rm cell}))$, where $f_{\rm coll}$ is the collapse fraction of minihalos above $M_c$ in that cell. The IGM temperature in the absence of astrophysical radiation is simply $T_{\rm K}^{\rm IGM}=T_0(1+\delta^{\rm cell}_{\rm IGM})^{2/3}$, where $T_0$ provided by {\tt RECFAST} \citep{Seager1999ApJ,Seager2000ApJS}.

\begin{figure}
    \centering
    \includegraphics[width=0.8\linewidth]{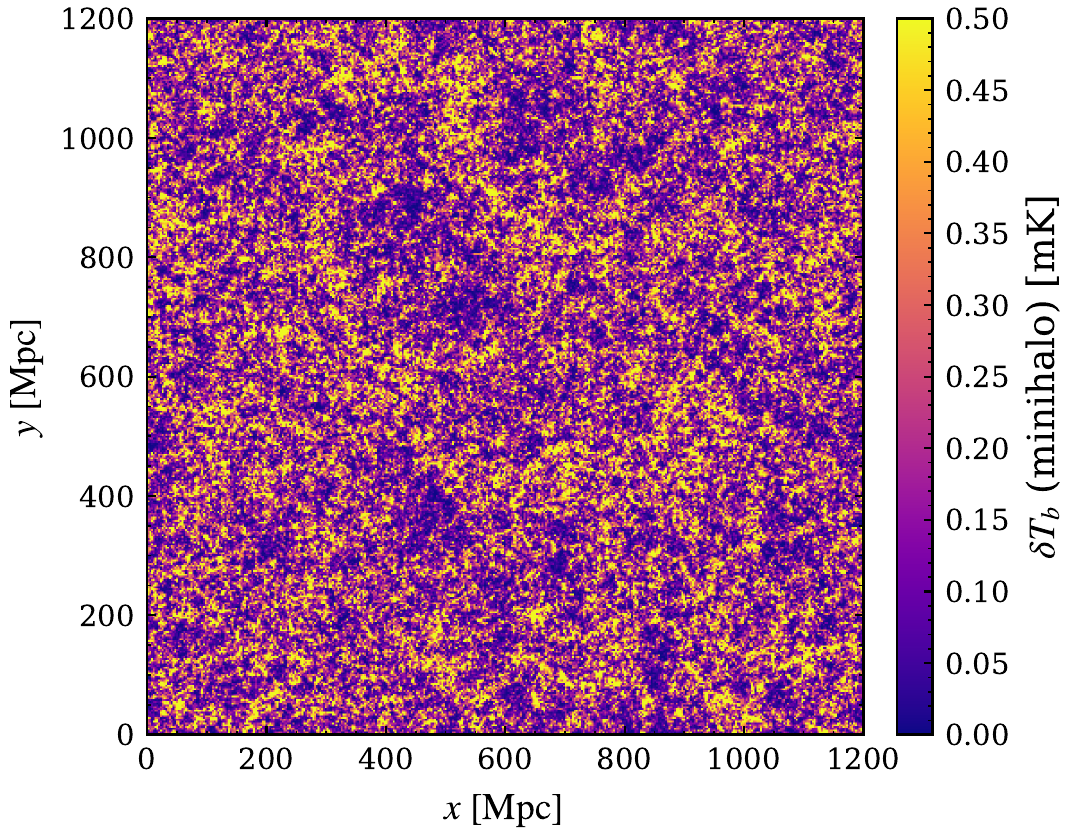}
     \includegraphics[width=0.8\linewidth]{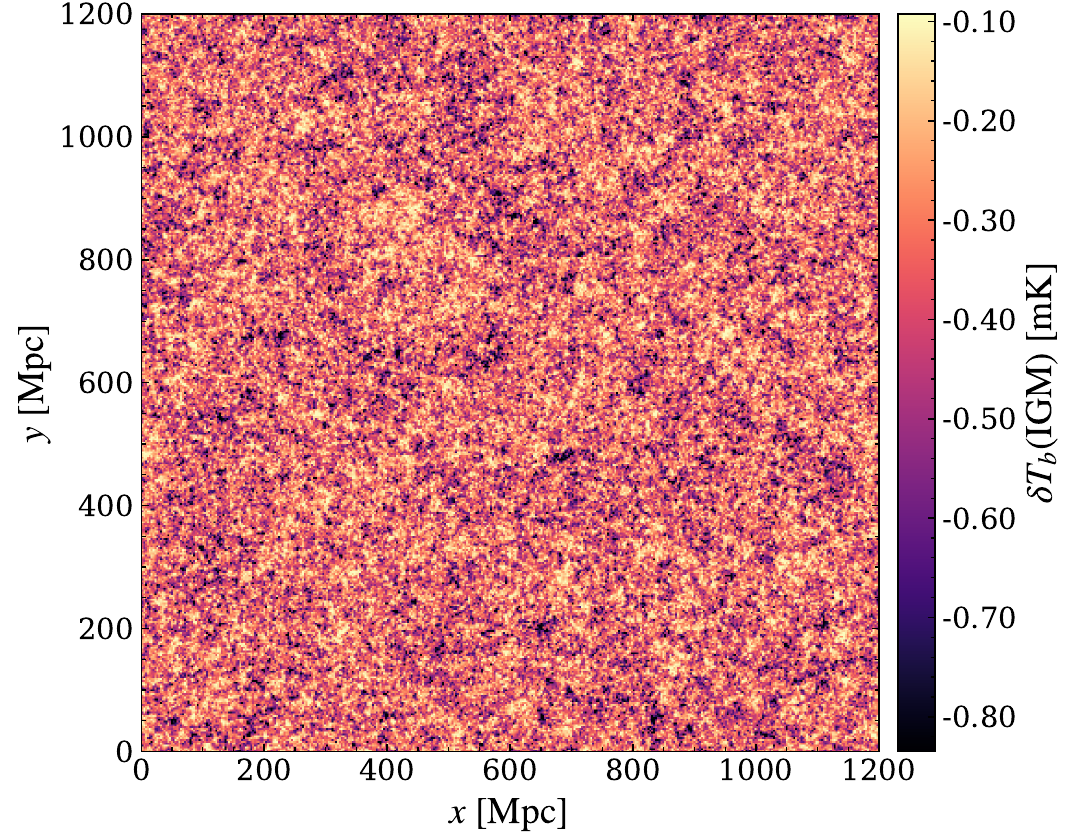}
    \caption{The fields of the 21 cm signal from minihalos (upper panel) and IGM (lower panel) at $z=20$, computed
according to the fields in Fig.~\ref{fig:vdb}. The slice thickness is
$3\,{\rm Mpc}$.
 }
    \label{fig:deltaT_b}
\end{figure}

\begin{figure}
    \centering
    \includegraphics[width=0.8\linewidth]{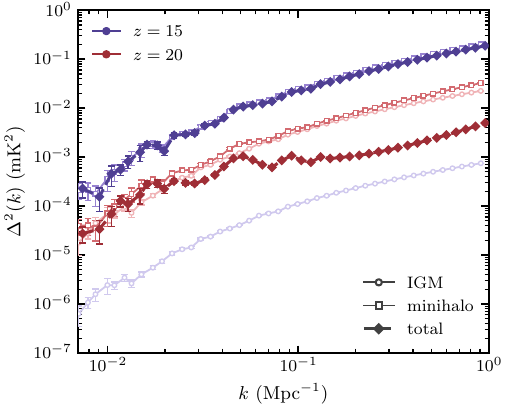}
    \includegraphics[width=0.8\linewidth]{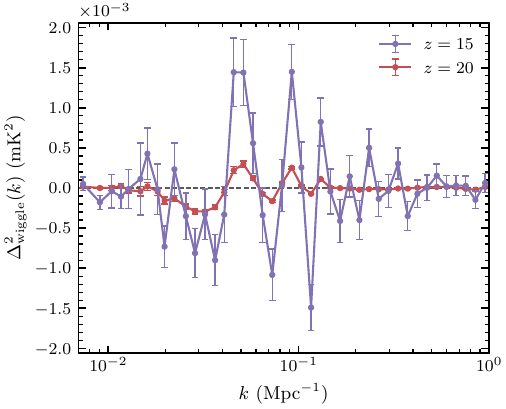}
    \caption{ 
{\it Top}: The power spectrum of the 21 cm signal from  minihalos and the IGM respectively, and that of the total signal, at $z= 15$ and $z= 20$. {\it Bottom}: the net wiggles in the power spectrum. The error bars correspond to the cosmic variance of the 21 cm power spectrum $\Delta^2(k)/\sqrt{N_{\rm m}(k)}$, where $N_{\rm m}(k)$ is the number of Fourier modes in that $k$-bin.
}
    \label{fig:ps_21cm}
\end{figure}

{\blue
Fig. \ref{fig:deltaT_b} exhibits 
}
slices of the 21 cm fields at $z= 20$ for minihalos and the IGM, respectively. For the three-dimensional boxes, at $z=20$, the mean signal is $0.23$ mK for minihalos, and $-0.35$ mK for the IGM. While at $z=15$, these two mean signals are $0.71$ mK and $-0.07$ mK respectively. The power spectra of these respective 21 cm signal and the total signal fields are shown in the top panel of  Fig. \ref{fig:ps_21cm}. 
We note that at $z=20$ the power spectrum of the total signal is even smaller than that of the individual components, i.e. minihalos or the IGM. This is because the total power spectrum includes a cross-correlation term between the minihalo emission and the IGM absorption fields. This term arises because minihalos form preferentially in overdense regions where the IGM density is also higher. Moreover, the 21 cm absorption of the IGM is also stronger in these overdense regions, where the collision coupling works more efficiently. 
Given that $\delta T_{\text{b,MH}}^{\rm cell} > 0$ (emission) and $\delta T_{\text{b,IGM}}^{\rm cell} < 0$ (absorption)
in the late Dark Ages, this cross-correlation is expected to be negative, thereby partially suppressing the total power spectrum amplitude. This is actually good for detecting the VAO wiggles. The VAO wiggles are primarily imprinted by the modulation of the minihalo abundance by the large-scale relative velocity field, whereas the IGM response to these velocities is negligible.  Consequently, the \textit{shape} and \textit{phase} of the VAO features remain dominated by the minihalo contribution, but the overall baseline amplitude is suppressed by the cross-correlation term. Therefore the VAO wiggles become more apparent. This ensures that the detectability forecasts presented in Sec. \ref{sec:detectability}, which focus on the significance of the wiggle features, remain robust. Same to \citet{Munoz2019}, we fit the 21 cm power spectrum by fifth-order polynomial, then remove the fitted smooth component to get the net VAO wiggles. They are plotted in the bottom panel of Fig. \ref{fig:ps_21cm}.

\section{The detectability}\label{sec:detectability}

We then estimate the detectability of the net VAO wiggles in the 21 cm power spectrum.
We adopt the same interferometric lunar-array configuration as the first line of the second item in Tab. 2 of \citet{Smith2025ApJ}: each sub-array is composed of $3\times 3$ cross-shaped dipoles; then, $4\times 4$ sub-arrays constitute a group; $2\times2$ groups, in turn, form one cluster; finally, the entire array consists of $12\times 12$ clusters and has 82944 dipoles in total, spread over $\sim$ 3.5 km. The layout of the array is similar to Fig. 1 of \citet{Smith2025ApJ}, except that the  $4\times 4$ dipoles in each sub-array (denoted by red squares in that figure) are replaced with $3\times 3$ dipoles. This configuration has signal-to-noise ratio (S/N) larger than the original {\it FarView} design \citep{Polidan2024,Burns_2026arXiv260116170B}.

We quantify the detectability of the net VAO wiggles by summing the S/N over the relevant $k$ bins,
\begin{align}
\left(\frac{S}{N}\right)_{\rm tot}
&=\left[\sum_i \left(\frac{S}{N}\right)_i^2\right]^{1/2}  \nonumber \\
&=\left[
\sum_i
\left(
\frac{\Delta^2_{\rm wiggles}(k_i)}
{\sigma_{\Delta^2 }(k_i)}
\right)^2
\right]^{1/2},
\label{eq:total_SN}
\end{align}
where $\Delta^2_{\rm wiggles}(k_i)$ is the net VAO wiggles amplitude in the
$i$th $k$ bin. The uncertainties $\sigma_{\Delta^2}$ combine the {\tt 21cmSense} forecast on observational uncertainties (include thermal noise and cosmic variance) with the uncertainties of the theoretical signal in quadrature,
\begin{equation}
\sigma_{\Delta^2}(k_i)
=
\sqrt{\sigma_{\rm obs}^2(k_i)
+
\sigma_{\rm theo}^2(k_i)}.
\end{equation}

\begin{figure}
    \centering
     \includegraphics[width=0.8\linewidth]{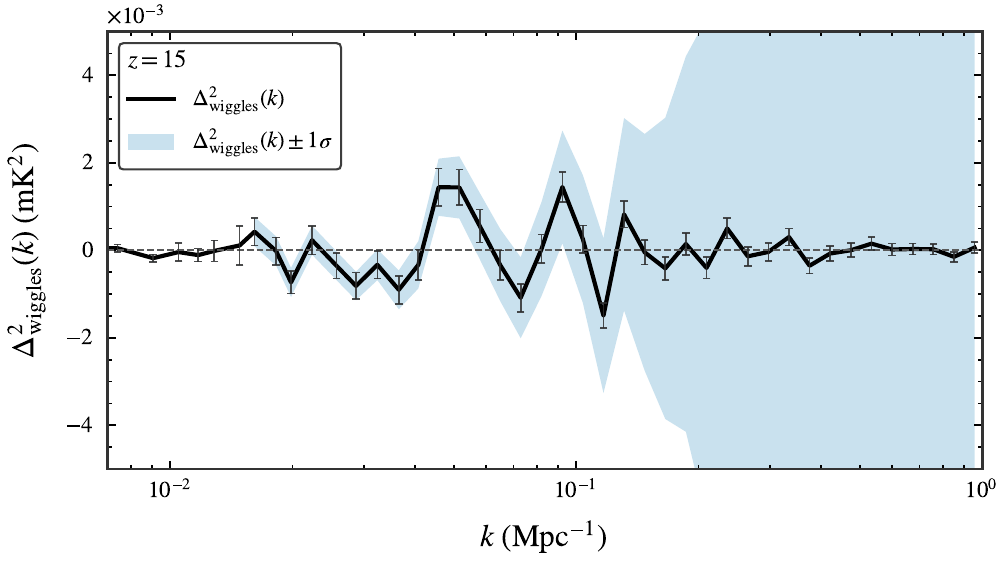}
 \includegraphics[width=0.8\linewidth]{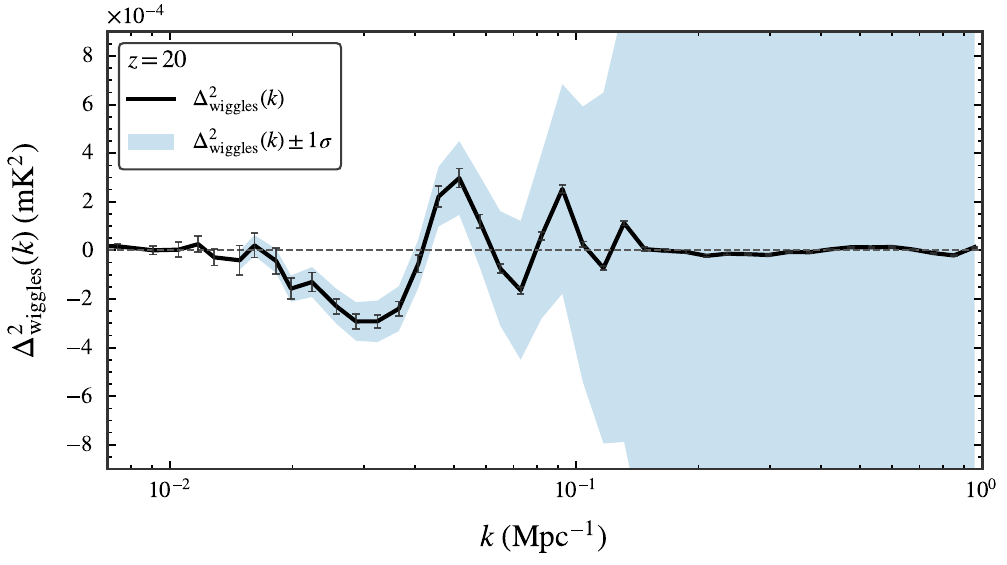}
 
    \caption{ 
{\it Top}: The net VAO wiggles from minihalos at $z=15$ and the predicted
observational uncertainties for an array of $12\times12$ clusters. The
integration time is set to approximately 24 lunar nights
($8064$ hr). 
{\it Bottom}: Same as the top panel, except that the redshift is $z=20$ and
the array has $22\times22$ clusters. The
integration time is set to 60 lunar nights ($20160$ hr). In each panel, the black error bars refer to cosmic variance of the theoretical signal, same as in Fig. \ref{fig:ps_21cm}; the blue shades indicate the observational uncertainties predicted by {\tt 21cmSense}. 
}
    \label{fig:ps_21cm_with_err}
\end{figure}

{\blue
The top panel of Fig.~\ref{fig:ps_21cm_with_err} shows
}
the forecast observational uncertainties on the net VAO wiggles at $z=15$ using this array configuration. The uncertainties are calculated using the package {\tt 21cmSense}
\citep{Murray2024,Pober2013,Pober2014,Parsons2012}, for an integration time of 24 lunar nights\footnote{Following the observing strategy considered in \citet{Smith2025ApJ}, we express the integration time in units of full lunar nights, with one lunar
night corresponding to 14 Earth days.} which are equal to 8064 hr. We adopt the optimistic foreground-subtraction model in {\tt 21cmSense}, which removes only the modes inside the primary-beam foreground wedge \citep{Pober_2016ApJ...819....8P}
. The wiggles are obviously detectable, with $\rm (S/N)_{\rm tot}\approx 4.7$.
At $z=20$, however, the signal is too weak to detect with this array configuration. Thus, we extend our array to $22\times 22$ clusters and show the predicted observational uncertainties for integration time of 60 lunar nights ($20160$ hr) in the bottom panel of Fig. \ref{fig:ps_21cm_with_err}.
With such an array configuration and  integration time, the net VAO wiggles are detectable as well, with $(S/N)_{\rm tot}\approx 7.4$.

\subsection{The VAO wiggles in cosmology beyond the standard $\Lambda$CDM model}

In the standard $\Lambda$CDM model, at $z=20$, the first peak (at $k=0.05$ Mpc$^{-1}$) of the net VAO wiggles from minihalos has amplitude $\sim 3.1\times 10^{-4}$ mK$^2$. 
For an interferometer to measure this scale at $z=20$, a baseline of $\sim$400 m is required, which necessitates the construction of a huge array. Moreover, since minihalos are not luminous, to achieve a statistically significant detection, it requires the dense baseline distribution formed by orders of $10^4$--$10^5$ antennas, as we have just demonstrated. 

{\blue 
There are, however, many beyond standard $\Lambda$CDM models in which the amplitude of small-scale matter power spectrum is much larger than in the standard $\Lambda$CDM model, but still consistent with the CMB observations. Examples include a running spectral index for primordial power spectrum \citep{Minoda2023}; a broad enhancement followed by a DM microphysical cutoff \citep{Inman2023}; a bump or tilt
on dwarf galaxy scales \citep{Skorikov2025}; and a localized Gaussian form power excess with a possible cutoff at still smaller scales
\citep{Nadler2025}.
\citet{Cielo_2025JCAP...04..007C} proposed a model for which, compared with the standard $\Lambda$CDM model, the primordial scalar power spectrum is enhanced on small scales by sudden phase transitions during inflation. There is a characteristic wavenumber $k_{\rm trans}$, around which the power spectrum first increases as $\propto k^6$; reaching a peak; then drops gradually. It forms a ``bump'' on the power spectrum as exhibited in Fig. 1 of \citet{Zhu_2026ApJ..1001...25Z}. This model has the advantage that, only the halos in a certain mass range is boosted, see Fig. 2 of \citet{Zhu_2026ApJ..1001...25Z}. By choosing  appropriate $k_{\rm trans}$,  the number of minihalos is boosted, but massive halos that have the potential to host galaxies are unchanged or just minorly changed. \citet{Zhu_2026ApJ..1001...25Z} found that in this model, the constraints on reionization history are not broken because the boost  on ionizing photons sinks (minihalos) balances the boost on ionizing photons sources. However, the morphologies of the ionization and the 21 cm fields change a lot, resulting in different bubble size distribution and 21 cm power spectrum.

In all these models, because the matter power spectrum is only enhanced in a certain $k$ range, compared with the standard $\Lambda$CDM model, minihalos could be orders of magnitude more numerous, while the number of massive halos capable of hosting luminous objects remains almost unaffected. The resulting enhancement of the 21 cm signal from minihalos provides an opportunity to constrain these non-standard small-scale
power spectra through their VAO features. For the quantitative
illustration below, we adopt the sudden-transition model of
\citet{Cielo_2025JCAP...04..007C}.
}

\begin{figure}
    \centering
\includegraphics[width=0.6\linewidth]{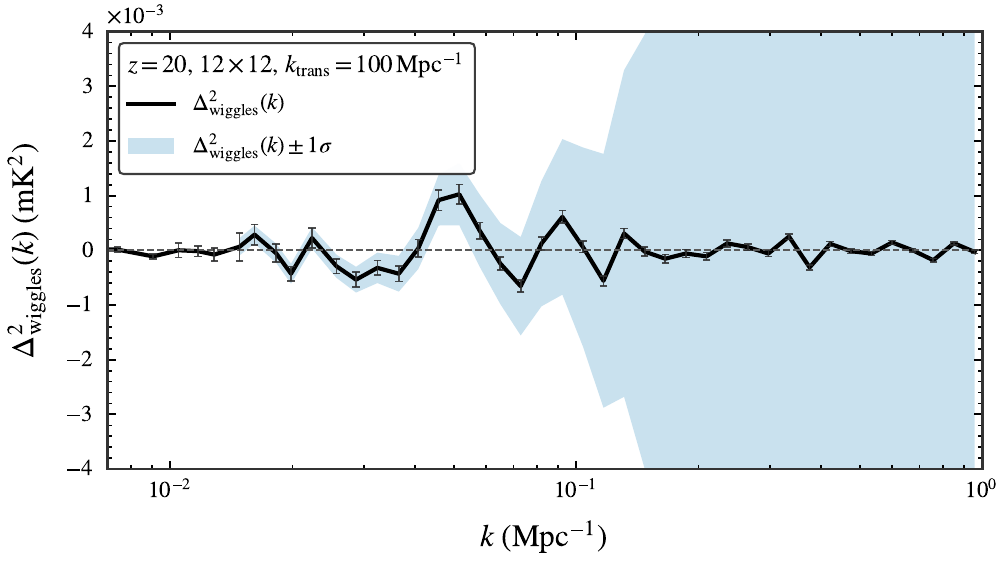}    
\includegraphics[width=0.6\linewidth]{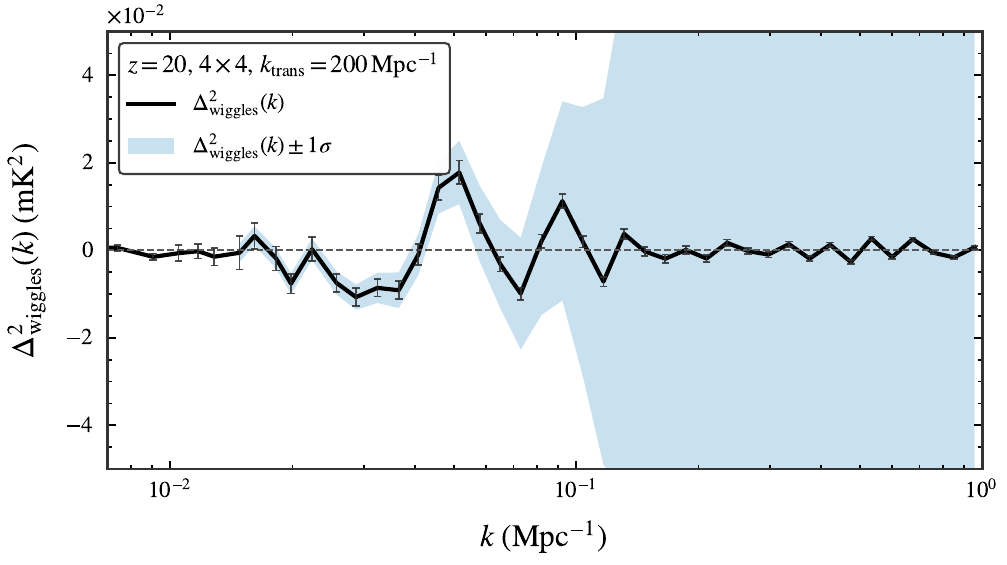}
\includegraphics[width=0.6\linewidth]{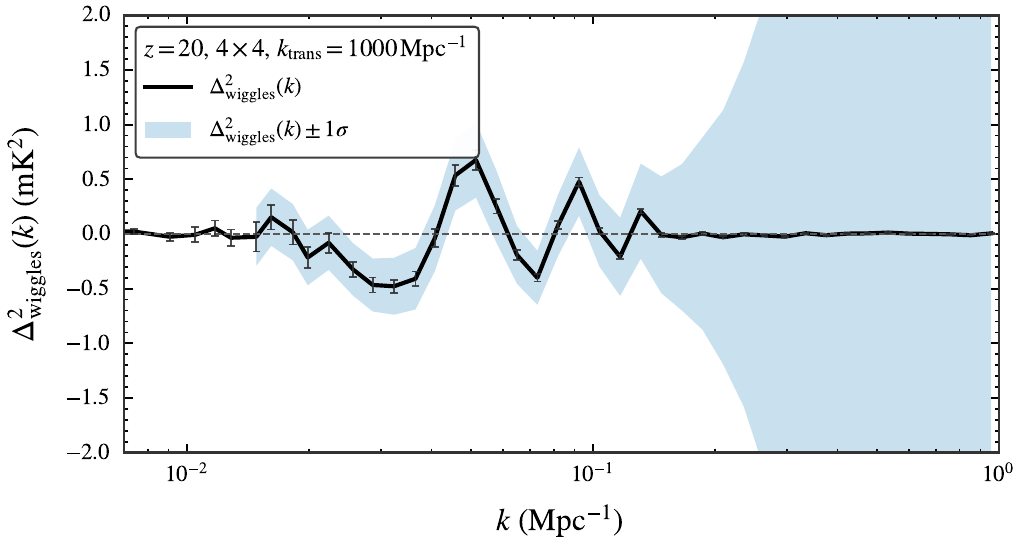}
    \caption{The net 21 cm VAO wiggles from minihalos at $z=20$ for the models of
\citet{Cielo_2025JCAP...04..007C} with $k_{\rm trans}=100~{\rm Mpc}^{-1}$
({\it top}), $200~{\rm Mpc}^{-1}$ ({\it middle}), and
$1000~{\rm Mpc}^{-1}$ ({\it bottom}), respectively. We also show the predicted
observational uncertainties for $12\times12$ clusters ({\it top}) and
$4\times4$ clusters ({\it middle}) assuming an integration time of
60 lunar nights ($20160$ hr), and for $4\times4$ clusters ({\it bottom}) assuming an
integration time of 6 lunar nights ($2016$ hr). In each panel, the black error bars
and blue shades denote the same types of uncertainties as in Fig.~\ref{fig:ps_21cm_with_err}.
The total S/N for the three models are
$4.6$, $6.4$, and $4.9$, respectively. 
}
    \label{fig:no_std_wiggle}
\end{figure}

In Fig. \ref{fig:no_std_wiggle}, we plot the net VAO wiggles in the 21 cm power spectrum from minihalos at $z=20$ with a transition wavenumber of $k_{\rm trans}=100$, $200$ and $1000$ Mpc$^{-1}$ respectively, along with predicted uncertainties for different array configuration and/or integration time. Other parameters are set the same as \citet{Zhu_2026ApJ..1001...25Z}. 
{\blue The figure exhibits }
that in such a model, the VAO wiggles become much larger. For $k_{\rm trans} = 100$ Mpc$^{-1}$, at $z=20$, the VAO signal improves by roughly an order of magnitude than in the $\Lambda$CDM model. When the transition wavenumber is pushed to 1000 Mpc$^{-1}$, the signal is already on the order of several mK$^2$. As a result, the wiggles can be detected with arrays consisting of fewer clusters of antennas. Under the $\Lambda$CDM model, $22 \times 22$ clusters and 60 lunar nights ($20160$ hr) integration time are required for detection at $z=20$. In this model, $12 \times 12$ clusters would suffice for the more conservative case of $k_{\rm trans}=100$ Mpc$^{-1}$. When the signal is greatly enhanced in models of a larger $k_{\rm trans}=1000$ Mpc$^{-1}$, only $4\times4$ clusters with 6 lunar nights ($2016$ hr) of integration are needed---required observational resources are greatly reduced.

\section{The contamination from the IGM}\label{sec:contamination}

{\blue
The IGM produces 21 cm signal whose absolute amplitude and power spectrum could be much larger than those of the minihalos. 
If the IGM temperature is  also influenced by the DM-baryon relative motion, then its power spectrum would also exhibit oscillatory features and may contaminate the VAO features of minihalos.
However, if its temperature is not or just slightly influenced by the relative motion, then its power spectrum will always be smooth and will not be confused with the VAO features of minihalos. In the above calculations we have ignored this possibility. In this section, we will investigate whether the IGM also has oscillatory features and whether it would contaminate the minihalso by using a hybrid of high-resolution small-box hydrodynamic simulations and large-scale semi-numerical simulations.
}

\subsection{\blue The small-box high-resolution hydrodynamical simulations}

We modify the {\tt Gadget-4} code \citep{Springel_2001NewA....6...79S,Springel2005MNRAS,Springel2021MNRAS} to include cooling and heating \citep{Xu2021ApJ}. We perform hydrodynamical simulations with a 3 Mpc box size and
$2\times1024^3$ particles. The mass resolutions of DM and gas particles are
approximately $8.4\times10^2\,M_\odot$ and $1.6\times10^2\,M_\odot$,
respectively. Such a
{\blue hydrodynamical simulation box}
mimics one cell in our cosmological box with $L_{\rm box}=1200\,{\rm Mpc}$. We then add different values of bulk DM-baryon relative streaming velocity $v_{\rm db}^{\rm cell}$ to the 
{\blue hydrodynamical simulation box}
to study its impacts on the IGM. 
We use the Friends-of-Friends (FoF) algorithm \citep{Davis1985ApJ} with linking length $b_{\rm lin}=0.2$ to find all DM particles in halos. Then, we remove all gas particles within the virial radius of each halo before calculating the net contribution to the mean 21 cm brightness temperature from the IGM.

\begin{figure}
    \centering
    \includegraphics[width=0.8\linewidth]{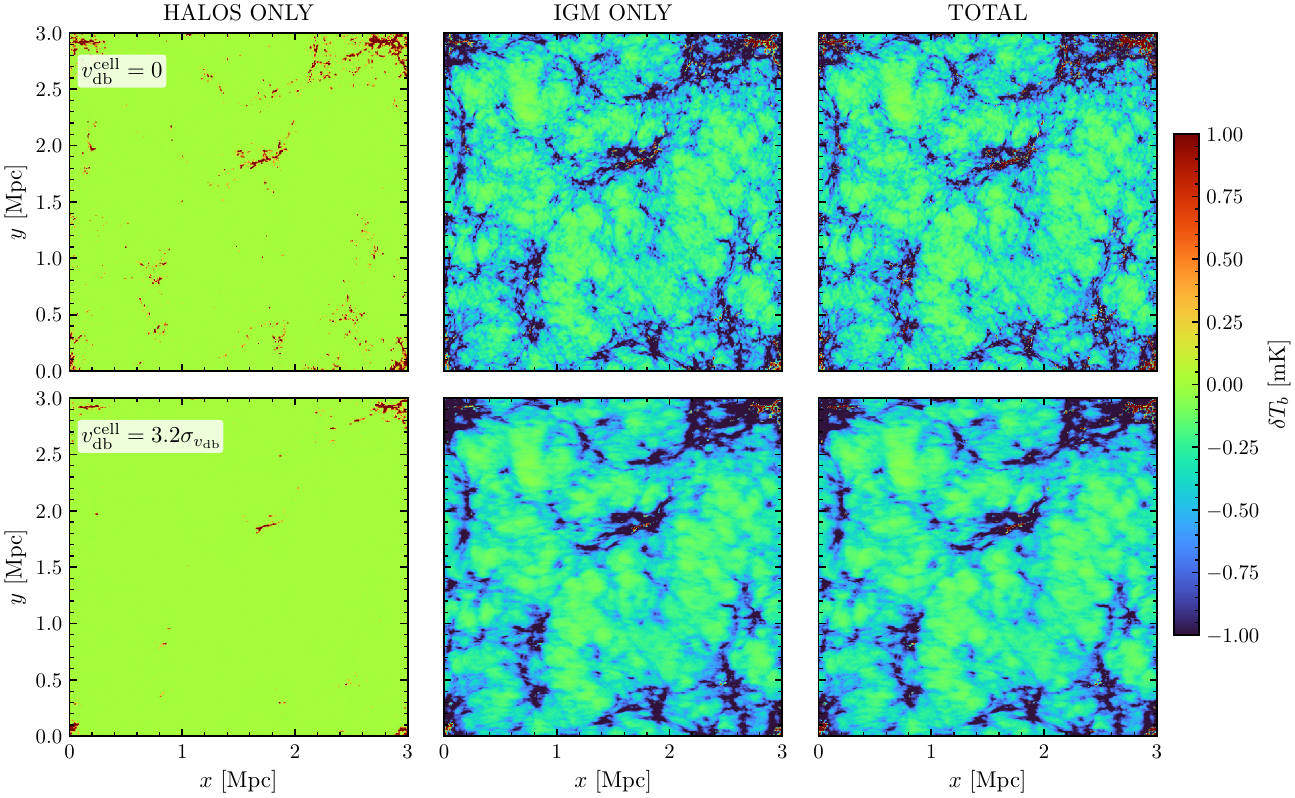}
    \caption{\blue {\it Left to right:} The 21 cm brightness temperature slices from minihalos,
the IGM after FoF halo removal, and their total signal at $z=20$ in a 3 Mpc
hydrodynamical simulation box. 
{\it Top and bottom:} Results for $v_{\rm db}^{\rm cell}=0$ and
$v_{\rm db}^{\rm cell}=3.2\sigma_{v_{\rm db}}$, respectively. The FoF linking
length is $b_{\rm lin}=0.20$, and the color scale shows $\delta T_b$ in mK.  
For all panels, the thickness of the slices is $188\,{\rm kpc}$.
 }
    \label{fig:3Mpc_slice}
\end{figure}

{\blue
Fig. \ref{fig:3Mpc_slice} shows the 21 cm signal slices in our hydrodynamical simulation box, for bulk velocities of zero and $\sim 3\sigma_{v_{\rm db}}$ respectively. 
Results for minihalos and the IGM, and their sum, are displayed separately.
The figure clearly indicates that,}
the bulk velocity changes the 21 cm signal of the minihalos significantly, while just very slightly influences that of the IGM.
We note that even in the IGM only panel, there are still some bright emission spots.
We check that they are hot and dense gas particles in the filaments outside virial radius of any minihalos.
Although we present the 21 cm signal from minihalos in Fig. \ref{fig:3Mpc_slice},
we also check that for these minihalos, the gas temperature drops to roughly half of the virial temperature at the virial radius; and the inner gas density is several factors lower than the TIS model. Therefore the 21 cm signal of individual minihalos in our simulations is smaller than the TIS model. It is not clear whether this is because our simulations suffer the resolution problem (many halos just have several tens to several hundreds gas particles), or because the TIS model is too optimistic for the 21 cm emission. Nevertheless,  we still adopt the TIS model to calculate the 21 cm emission of minihalos. The minihalos panel in Fig. \ref{fig:3Mpc_slice} is just used to intuitively show the DM-baryon relative motion effects.

\begin{figure}
    \centering
 \includegraphics[width=0.8\linewidth]{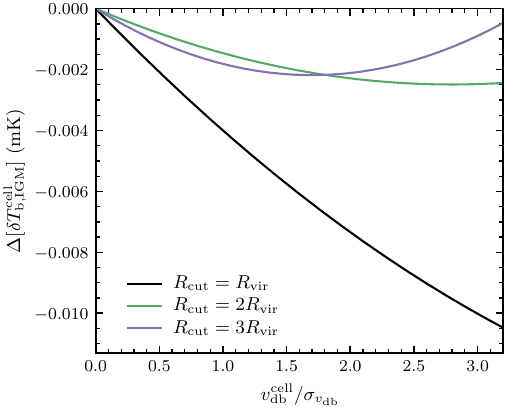}
    \caption{\blue 
The change in the mean 21 cm brightness temperature of the 3 Mpc box induced by the bulk DM-baryon relative motion, as a function of the bulk velocity, at $z=20$. For displaying purpose, the $x$-axis is normalized by $\sigma_{v_{\rm db}}$ at $z=20$. We remove the gas particles within a certain radius of  halos found by the FoF algorithm, as indicated in the legend. 
}
    \label{fig:dT_vs_v}
\end{figure}

{\blue The difference between the mean 21 cm signal of the IGM in a 3 Mpc hydrodynamical simulation box} with and without the bulk relative streaming velocity, 
\begin{equation}
\Delta [\delta T_{\rm b,IGM}^{\rm cell}](v_{\rm db}^{\rm cell})=\delta T_{\rm b,IGM}^{\rm cell}(v_{\rm db}^{\rm cell})-\delta T_{\rm b,IGM}^{\rm cell}(v_{\rm db}^{\rm cell}=0),
\label{eq:Delta_Tb_IGM}
\end{equation}
as a function of $v_{\rm db}^{\rm cell}$, is plotted in Fig. \ref{fig:dT_vs_v} at $z=20$. 
{\blue The} magnitude of $\Delta [\delta T^{\rm cell}_{\rm b,IGM}]$ appears to increase with $v_{\rm db}^{\rm cell}$, but by less than $\sim 0.01$ mK when the bulk velocity $\lesssim 3\sigma_{v_{\rm db}}$. 
To further investigate the positions of the gas particles contributing to $\Delta [\delta T_{\rm b,IGM}^{\rm cell}]$, we also remove the gas particles within $2R_{\rm vir}$ and $3R_{\rm vir}$ respectively. The  resulting $\Delta [\delta T_{\rm b,IGM}^{\rm cell}]$ is shown  in Fig. \ref{fig:dT_vs_v}  as well. 
We see that in this case, the magnitude {\blue becomes} $\lesssim 0.002$ mK. It means that, most of the shock-heated IGM gas particles are actually located in the vicinity of halos. {\blue This is consistent with the discoveries in \citet{Oleary2012ApJ}.}
By inspecting the simulation outputs, we also find that some shock-heated gas particles {\blue are} located in filaments even beyond \(3R_{\rm vir}\) from the center of FoF halos.
This indicates that the separation between the immediate
halo environment and the diffuse IGM is not perfectly sharp, but it does not
affect our conclusion that, for the 21 cm signal, the IGM response to the bulk relative velocity is
negligible compared with the minihalos signal.

\subsection{\blue The hybrid method and large-scale VAO features of the IGM}

\begin{figure*}
    \centering
      \includegraphics[width=0.45\linewidth]{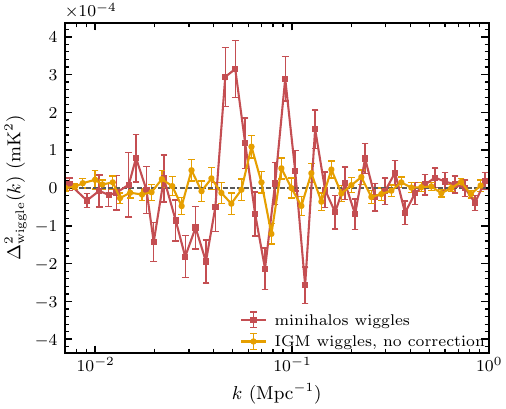}
         \includegraphics[width=0.45\linewidth]{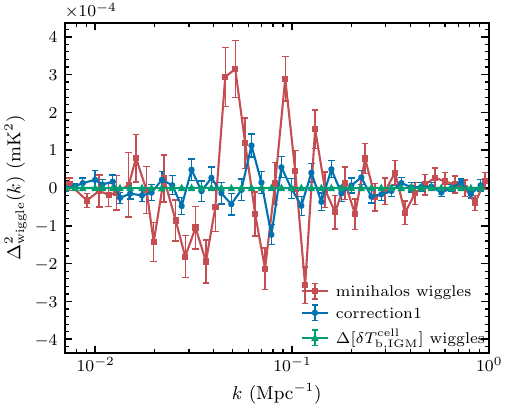}
      \includegraphics[width=0.45\linewidth]{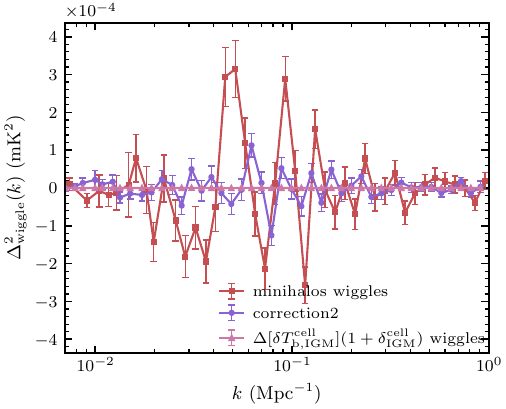}
 \includegraphics[width=0.45\linewidth]{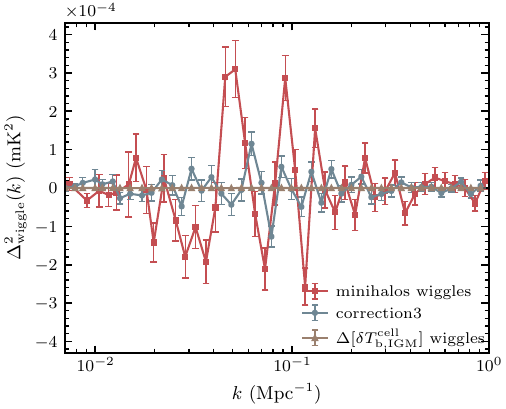} 
    \caption{\blue 
{\it Top left}: The net wiggles component of the IGM power spectrum without adding any velocity-induced correction, compared with the minihalos wiggles at the same redshift.
{\it Top right}:
Same as the top left panel, however the correction of hypothesis 1 is added to the IGM. The wiggles of the correction field itself is also displayed.
{\it Bottom left:} Same as the top right panel, except that for the hypothesis 2.
{\it Bottom right:} Same as the top right panel, except that for the hypothesis 3.
All panels are for $z=20$.
In each panel, the net wiggles components are obtained by subtracting a fifth-order polynomial smooth fit from the corresponding power spectra.
The last three panels are almost the same, implying that the relative motion generates negligible wiggles on the 21 cm signal of the IGM.
}
    \label{fig:ps_offset}
\end{figure*}

{\blue
Although the small-box high-resolution hydrodynamical simulations reveal that the 21 cm signal of the IGM is quite less sensitive to the relative motion, they cannot directly provide the estimate of the VAO wiggles signal for the IGM, because the wiggles are large-scale features.  On the other hand, however, it is quite impractical to run hydrodynamical simulations with box size as large as our cosmological box, while at the same time can resolve the smallest minihalos. A popular method for solving this probelm in literature is the hybrid method (e.g. \citealt{Visbal2012Nature,Fialkov2013MNRAS,McQuinnOLeary2012ApJ,Fialkov2014IJMPD,Park_2025NatAs...9.1723P,Sikder_2026ApJ..1000L..50S}). 
Using many small-box hydrodynamical simulations with same  box size equal to the cell size of the large-scale cosmological box, however with different bulk relative velocities and background overdensities, to construct a table of 21 cm signals as a function of velocity and overdensity. Then, for each cell in the large-scale box, its 21 cm signal is interpolated based on this table.

In our work, using the hybrid method we should add a velocity- and overdensity-dependent correction term
\begin{equation}
\Delta[\delta T_{\rm b,IGM}^{\rm cell}]
(v_{\rm db}^{\rm cell},\delta_{\rm cell})
=
\delta T_{\rm b,IGM}^{\rm cell}
(v_{\rm db}^{\rm cell},\delta_{\rm cell})
-
\delta T_{\rm b,IGM}^{\rm cell}
(0,\delta_{\rm cell}),
\label{eq:igm_velocity_density_response}
\end{equation}
to the 21 cm field generated from Eq. (\ref{eq:T_b_IGM}), to account for the relative motion effects on the IGM's 21 cm signal.

However, all our small-box hydrodynamical simulations have zero background overdensity, in principle we do not know how their 21 cm signal depends on the background overdensity. We analyze the cross-correlation between our large-scale density field and relative velocity field, and find that, the cross-correlation coefficient is at $\mathcal{O}(10^{-4})$ level. 
So these two fields are almost independent of each other \citep{Cain2020ApJ}. 
We therefore reasonably hypothesize that, the behavior of $\Delta[\delta T_{\rm b,IGM}^{\rm cell}](v^{\rm cell}_{\rm db},\delta_{\rm cell})$ as a function of background overdensity $\delta_{\rm cell}$, is independent of relative motion. That is 
\begin{equation}
\Delta[\delta T_{\rm b,IGM}^{\rm cell}](v^{\rm cell}_{\rm db},\delta_{\rm cell})\approx 
\Delta[\delta T_{\rm b,IGM}^{\rm cell}(v^{\rm cell}_{\rm db},0)]\times \mathcal{F}(\delta_{\rm cell}). 
\end{equation}
We try three hypotheses: (1) $\mathcal{F}=1$; (2) $\mathcal{F}=1+\delta_{\rm IGM}^{\rm cell}$; and (3) $\mathcal{F}=\delta T_{\rm b,IGM}(\delta_{\rm IGM}^{\rm cell})/\delta T_{\rm b,IGM}(0)$, for which $\delta T_{\rm b,IGM}$ is the 21 cm brightness temperature of the IGM in the absence of relative motion, expressed in Eq. (\ref{eq:T_b_IGM}).  

For all of these three hypotheses, we find that the oscillatory features of the IGM arising from the relative motion are almost negligible compared with minihalos. 
}
In Fig. \ref{fig:ps_offset} we plot the net wiggles in the 21 cm power spectrum of the IGM {\blue at z=20}, in the absence ({\blue top left panel}) and presence ({\blue top right panel for hypothesis 1, bottom left panel for hypothesis 2 and bottom right panel for hypothesis 3}) of relative motion correction respectively.  For comparison we also plot the VAO wiggles from the minihalos (same to the bottom panel in Fig. \ref{fig:ps_21cm_with_err}). We find that, even in the absence of DM-baryon relative motion correction, the 21 cm power spectrum of the IGM has its own wiggles. The locations of these wiggles are different from the VAO wiggles, they are actually generated by the baryon acoustic oscillations (BAO) in the density field \citep{Zhang2024,Cruz_2026arXiv260709846C}. 
Such BAO wiggles may confuse the VAO signal, fortunately, their  amplitude is  obviously smaller than the VAO wiggles of minihalos. 
{\blue Moreover, even after adding the bulk motion-induced temperature correction to the IGM, the wiggles in its 21 cm power spectrum change very little.} It means that the effects of the DM-baryon relative motion on the IGM is negligible compared with the IGM's BAO wiggles and the VAO wiggles of minihalos. To further confirm this, we also plot the power spectra of 
{\blue 
the correction term  
{\blue for the three hypotheses }
respectively in Fig. \ref{fig:ps_offset}.
}
Indeed, they are negligible compared with the VAO wiggles of minihalos. 
{\blue
We conclude that the contamination from the IGM can be safely ignored in detecting the VAO features of minihalos in the late Dark Ages.
}

{\blue
We clarify that, when comparing the VAO features with the BAO features, the situation is different for the Dark Ages and the Cosmic Dawn, and different for minihalos and the IGM. 

In the Dark Ages, astrophysical luminous objects have not yet formed, so the amplitudes of the VAO and BAO wiggles do not depend on astrophysical parameters. The 21 cm signal from minihalos has both the VAO wiggles and BAO wiggles. This is because the BAO component is inherited from acoustic features in the density field, while the abundance of minihalos is modulated by their environment densities.  In contrast, the VAO component considered here arises because the coherent DM-baryon relative velocity field modulates gas content of low-mass minihalos. Its amplitude is therefore sensitive to the velocity-dependent mechanism that prevent gas from being accreted into minihalos. That is to say, the amplitude of the VAO wiggles is sensitive to the lower limit of minihalos; while the amplitude of the BAO wiggles depends on the cumulative gas content in all minihalos. So the shape of minihalo mass function is crucial in determining their relative amplitude, which of course depends on cosmology parameters. Regarding the IGM in the Dark Ages, according to our hypotheses, its VAO wiggles are much weaker than the BAO wiggles.  The amplitude of BAO wiggles replies on the primordial matter power spectrum and the growth of density perturbations, which  is further dependent on cosmology parameters.

In the Cosmic Dawn, the IGM has both the VAO and BAO wiggles, and their relative amplitude evolves along with the formation of first stars \citep{Zhang2024,Cruz_2026arXiv260709846C}. At the beginning of the Cosmic Dawn, a small amount of the first stars just form, the IGM only exhibits BAO wiggles and its VAO wiggles are negligible, like in the Dark Ages. As the first stars form in large numbers and the IGM is significantly influenced by their Ly$\alpha$ and X-ray radiation, the VAO wiggles become much higher than the BAO wiggles. Later on, when the Ly$\alpha$ scattering and X-ray heating become saturated, the VAO wiggles disappear but the BAO wiggles always exist because it is directly from the coherent structures in the density field. The formation of first stars is regulated by both astrophysical parameters and cosmological parameters, so in the Cosmic Dawn, both the VAO and BAO wiggles and their relative amplitude depend on these parameters.

In this work, we limit our investigations to the Dark Ages. We clarify that our results about the amplitudes of the VAO and BAO wiggles, and their comparison, apply only to the cosmology and minihalo model adopted in this paper. Although the characteristic acoustic shape of the VAO template is comparatively robust \citep{Munoz2019}, the relative BAO and VAO amplitude varies with model parameters.
}

\section{Discussion}\label{sec:discussion}

The VAO features from minihalos evolve rapidly and the lightcone effect must be significant. For example, at $z=30$, the amplitude of the first peak at $k=0.05$ Mpc$^{-1}$ is $9.7 \times 10^{-7}$ mK$^2$; at $z=20$ it is $3.1\times 10^{-4}$ mK$^2$; and at $z=15$ it is $1.4\times 10^{-3}$ mK$^2$.  We have ignored the lightcone effect and may overestimate the ${\rm (S/N)_{tot}}$. In such a case, the multi-frequency angular power spectrum is more appropriate than the spherically-averaged 3D power spectrum \citep{Du2026}.

Another challenge is the foreground radiation from our Milky Way and extragalactic sources. It can be $\sim 5$ orders of magnitude higher than the 21 cm signal from the Dark Ages. Although it is spectrally smooth and many foreground removal algorithms take advantages of this feature,  
the chromatic response of the interferometer mixes angular foreground structures into frequency structures, contaminating the signal in a wedge-like region in the $(k_\perp,k_\parallel)$ space, where $k_\perp$/$k_\parallel$ is the wavenumber  perpendicular/parallel to the line of sight (e.g. \citealt{Datta2010ApJ,Morales_2012ApJ...752..137M,Parsons2012,Pober2014,Liu_2014PhRvD..90b3018L,Liu_2014PhRvD..90b3019L,Jensen_2016MNRAS.456...66J,Gagnon2021MNRAS}).
The wedge border is set either by the horizon (horizon-limit) or by the primary beam (primary beam-limit).
Modes outside this region are relatively cleaner. The most straightforward method is to only use these clean modes, and this is the foreground-avoidance.
However, the impact of the wedge becomes more severe at higher redshift because the slope of its upper boundary increases with redshift, reducing the number of accessible modes outside the wedge. Therefore, the foreground contamination presents a greater challenge for the Dark Ages than the epoch of reionization.

\citet{Pober2025ApJ} quantified this effect for the Dark Ages experiments and found that, using the horizon-limit wedge, foreground-avoidance typically reduces the power spectrum detection significance by about an order of magnitude. For their fiducial array, at $z=30$, the nominal no-foreground detection significance for the 21 cm power spectrum decreases from $10\sigma$ to $\sim1.1\sigma$ after the modes in the foreground wedge are discarded. Similarly, \citet{Bull2024RSPTA} found that, excluding the modes in the worst case foreground wedge degrades the constraints on the non-Gaussianity in the bispectrum by a factor of $\gtrsim 5- 100$.

\citet{Ghosh2026MNRAS} performed the first end-to-end simulation pipeline for the Dark-ages EXplorer. Their mock data includes the 21 cm signal from the Dark Ages, the foreground, thermal noise, and the lunar topography. They found that: using the polynomial fitting and spectral tapering, the 21 cm signal from $z\sim 42$ can be recovered, although the antenna-position and lunar-topography errors can bias the recovered power spectrum by $\sim10-30\%$; but the signal from $z\sim 148$ is still difficult to access. Their findings boost our confidence in detecting the 21 cm signal from the late Dark Ages ($z\sim 20$).

Since the VAO wiggles are large-scale features, foreground-removal is necessary \citep{Zhang2024}. We note that in  \citet{Ghosh2026MNRAS}, even for the $z\sim 42$ signal, the modes with $k_\parallel \lesssim 0.2$ Mpc$^{-1}$ are still contaminated by foreground. In this case, the multi-frequency angular power spectrum that only counts the correlation from modes perpendicular to the line of sight could be useful \citep{Du2026}, and probably more advanced foreground-removal algorithms are required. 

In this paper, we adopt the foreground--avoidance strategy. However, our primary beam is the synthetic beam of the sub-array, which is much smaller than a dipole. Therefore our foreground-wedge is actually much smaller than the horizon-limit and the dipole beam-limit. Our results are actually close to the perfect foreground-removal case, so the forecasts presented in this paper are considered as the optimistic case.

\section{Summary } \label{sec:summary}

In this paper, we investigated the VAO features on the 21 cm power spectrum of minihalos in the late Dark Ages ($z\sim 20$) and their detectability by low-frequency array at the farside of lunar surface.
During this epoch, if the IGM has never been influenced by the radiation from Pop III stars, its 21 cm signal is still in absorption, but much weaker due to the less efficient {\blue collisional} coupling compared with higher redshifts,
while the signal from minihalos is in emission. At $z\gtrsim 30$, the fluctuations amplitude of the IGM's 21 cm signal is  much larger than that of minihalos, but they become comparable at $z\sim 30 -20$. Generally, it is not easy to isolate the contribution from minihalos in the detected total 21 cm signal. However, the relic supersonic DM-baryon relative streaming velocities suppress the gas accretion onto minihalos and modulate their 21 cm signal, finally 
{\blue leaving} the VAO features in the power spectrum. 
{\blue
On the other hand, the 21 cm signal from the IGM is just slightly influenced by such relative velocities and exhibit negligible VAO features compared with that from minihalos.} 
This provides us with a chance to probe minihalos in the late Dark Ages. We calculated the VAO wiggles and estimated the observational uncertainties, we found that:
\begin{enumerate}
    \item In the standard $\Lambda$CDM cosmology, the first peak of the net VAO wiggles at $k=0.05$ Mpc$^{-1}$ has amplitude $\sim 1.4\times 10^{-3}$ mK$^2$ at $z\sim 15$, and  $\sim 3.1\times 10^{-4}$ mK$^2$ at $z\sim 20$, respectively. 
    
    \item If the IGM has not yet been influenced by radiation from Pop III stars, an array with 82944 dipoles, like the one proposed in \citet{Smith2025ApJ}, can detect the VAO features from minihalos at $z\sim 15$ with 24 lunar nights
($8064$ hr). If the array can be extended by a factor of $\sim 3$, then it can detect the VAO features from $z\sim 20$ with 60 lunar nights ($20160$ hr).
    
    \item In some cosmology models, the small-scale matter power spectrum  is enhanced compared with the standard $\Lambda$CDM model.  In such models, minihalos can produce much stronger VAO wiggles that are detectable for a smaller array, providing us with a new chance for constraining these models.   
\end{enumerate}

Our investigation provides a new strategy for probing the small-scale nonlinear structures in the late Dark Ages by the lunar-based low-frequency array on the farside of the Moon in the future.

\begin{acknowledgments}
We thank the anonymous referee very much for the constructive suggestions that help to improve this manuscript.
This work is supported by the NSFC International (Regional) Cooperation and Exchange Project No. 12361141814, the National SKA Program of China Nos. 2020SKA0110402 \& 2020SKA0110401, the China's Space Origins Exploration Program Nos. GJ11010405 \& GJ11010401, the National Key R\&D Program of China No. 2022YFF0504300, and the Project Supported by the Specialized Research Fund for State Key Laboratory of Radio Astronomy and Technology. The work was carried out at National Supercomputer Center in Tianjin, and the calculations were performed on Tianhe new generation supercomputer.
\end{acknowledgments}

\bibliography{references}{}
\bibliographystyle{aasjournalv7}



\end{document}